\documentclass[copyright,creativecommons]{eptcs}
\providecommand{\event}{ACL2 Workshop 2025}
\usepackage{underscore,
  latexsym,pifont,amsmath,amstext,amssymb,amsfonts,makeidx,url, breakurl}
\title{A Formalization of Elementary Linear Algebra: Part I}
\author{David M.~Russinoff
\email{david@russinoff.com}
}
\def\titlerunning{A Formalization of Linear Algebra: Part I}
\def\authorrunning{D.M. Russinoff}

\begin{document}
\maketitle

\begin{abstract}
This is the first installment of an exposition of an ACL2 formalization of elementary linear algebra, focusing on aspects of the subject that apply to matrices
over an arbitrary commutative ring with identity, in anticipation of a future treatment of the characteristic polynomial of a matrix, which has entries in a
polyniomial ring.  The main contribution of this paper is a formal theory of the determinant, including its characterization as the unique alternating $n$-linear
function of the rows of an $n \times n$ matrix, multiplicativity of the determinant, and the correctness of cofactor expansion.
\end{abstract}

\section{Introduction}

This is the first installment of an exposition of an ACL2 formalization of elementary linear algebra, covering the basic algebra of matrices and the theory of 
determinants. Part II \cite{linear2}, also included in this workshop, addresses row reduction and its application to matrix invertibility and simultaneous systems
of linear equations.  Additional topics to be covered in future installments include vector spaces, linear transformations, polynomials, eigenvectors. and
diagonalization.

This ordering of topics departs from the typical syllabus of an introductory course in the subject. Most elementary linear algebra textbooks treat the solution of
simultaneous linear equations in the first chapter, perhaps to reassure the student of the practical utility of the theory.  Consequently, (since this process 
depends on the existence of a multiplicative inverse) the entries of a matrix are assumed at the outset to range over a field (often the real numbers) rather than
a more general commutative ring.  This assumption, however, is not required for the main results of matrix algebra or the properties of the determinant; in fact, there
are numerous applications for which it does not hold \cite{brown}.   Indeed, several chapters later, one finds that the theory of eigenvalues is based on the
fundamental notion of the characteristic polynomial of a matrix over a field $F$, which is properly defined as the determinant of a matrix with entries in the
polynomial ring $F[t]$.  In most cases, this problem is simply ignored \cite{kolman, kwak, roman}.  A rare exception is a comparatively rigorous treatment by
Hoffman and Kunze \cite{hoffman}, on which our formalization is partly based (and from which this author learned the subject as a college sophomore).
In Chapter 5, (anticipating the introduction of the characteristic polynomial) they define the determinant of a matrix over an arbitrary commutative ring with
unity and ask the reader to determine for himself which of the results of the preceding chapters, though stated and proved for matrices over a field, apply more
generally to commutative rings.

Neither of these strategies will serve our purpose.  Unconstrained by pedagogical considerations, we pursue a more principled development, separating those aspects
of the theory that are valid for general commutative rings from those that depend on the existence of a multiplicative inverse.  The former topics are treated
in this paper; the latter in Part II.  All supporting proof scripts reside in the shared ACL2 directory {\tt books/projects/linear/}.

In Section~\ref{ring}, we introduce the notion of an abstract commutative ring with unity by means of an encapsulated set of constrained functions
and associated theorems corresponding to the standard ring axioms.  Section~\ref{mat} covers the algebra of matrices and the transpose operator.  The main
contribution of this paper is a formal theory of determinants, based on the classical definition, which appeals to the properties of the symmetric group.
This consideration was a factor in our broader plan of a formalization of algebra beginning with the theory of finite groups \cite{groups1, groups2, groups3},
on which the present work critically depends.  In Section~\ref{det}, we define the determinant and derive its main properties, including its uniqueness as an
alternating $n$-linear function of the rows of an $n \times n$ matrix.  Multiplicativity is derived as a consequence of this result, which is further exploited in
Section~\ref{cofactors} to establish the correctness of cofactor expansion and the properties of the classical adjoint.  The proofs of uniqueness and its
consequences illustrate the use of encapsulation and functional instantiation as a substitute for higher order logical reasoning in the first order logic of ACL2.

Previous work on matrix algebra within the ACL2 community includes a formalization by Gamboa et al.~based on ACL2 arrays \cite{cowles}, another by Hendrix with
matrices defined as simple list structures \cite{hendrix}, and Kwan's proofs of correctness of several numerical algorithms \cite{kwan1, kwan2}.  Our matrix
representation scheme is essentially that of Hendrix (which was also adopted by Kwan), but since we require entries ranging over an abstract ring rather than the
{\tt acl2-number} type, ours is constructed independently.  Only the first of these cited references provides a general definition of the determinant, with no
proofs of its properties.  A variety of linear algebra formalizations have been based on other theorem provers \cite{lean, coq, coq2, hol4, isabelle}, but we are
not aware of any that has produced the full list of results reported above.

\section{Commutative Rings}\label{ring}

In {\tt ring.lisp}, the axioms of a commutative ring with unity are formalized by an encapsulation, partially displayed below:

\begin{small}
\begin{verbatim}
  (encapsulate (((rp *) => *)                   ;ring element recognizer
                ((r+ * *) => *) ((r* * *) => *) ;addition and multiplication
                ((r0) => *) ((r1) => *)         ;identities
                ((r- *) => *))                  ;additive inverse
    (local (defun rp (x) (rationalp x)))
    (local (defun r+ (x y) (+ x y)))
    (local (defun r* (x y) (* x y)))
    (local (defun r0 () 0))
    (local (defun r1 () 1))
    (local (defun r- (x) (- x)))
    ;; Closure:
    (defthm r+closed (implies (and (rp x) (rp y)) (rp (r+ x y))))
    (defthm r*closed (implies (and (rp x) (rp y)) (rp (r* x y))))
    ;; Commutativity:
    ...
  }
\end{verbatim}
\end{small}
This introduces six constrained functions: {\tt rp} is a predicate that recognizes an element of the ring; {\tt r+} and {\tt r*} are the binary addition and
multiplication operations; the constants {\tt (r0)} and {\tt (r1)} are the identity elements of these operations, respectively; and {\tt r-} is the unary addition
inverse.  Note that these functions are locally defined to be the corresponding functions pertaining to the rational numbers (an arbitrary choice---the recognizer
{\tt (integerp x)} would have worked just as well as {\tt (rationalp x)}).  The exported theorems (mostly omitted above) are the usual ring axioms: closure, 
commutativity, and associativity of both operations; properties of the identities and the additive inverse; and the distributive law.  Informally, we shall refer
to the ring {\tt R} that is characterized by these axioms, and elements of {\tt R} are sometimes called {\it scalars}.  When our intention is clear, we may abbreviate
{\tt (r0)} and {\tt (r1)} as 0 and 1, respectively.

The file also contains some trivially derived variants of the axioms, along with definitions of several functions pertaining to lists of ring elements and proofs of their basic properties:

\begin{itemize}
\item {\tt rlistp} is a predicate that recognizes a vector, i.e., a proper list of scalars, which we call an
{\it rlist};
\item {\tt rlistnp} recognizes an rlist of a specified length;
\item {\tt rlist0p} recognizes an rlist of which every member is {\tt (r0);}
\item {\tt rlistn0} returns an rlist of a specified length of which every member is {\tt (r0)};
\item {\tt rlist-sum} and {\tt rlist-prod} compute the sum and product, respectively, of the members of an rlist;
\item {\tt rlist-scalar-mul} multiplies each member of an rlist by a given scalar and returns a list of the products;
  \item {\tt rdot} computes the {\it dot product} of two rlists of the same length, i.e., the sum of the products of corresponding members;
\item {\tt rdot-list} returns the list of dot products of an rlist with the members of a list of rlists.
\end{itemize}
The reader may anticipate that a function name containing the character {\tt r}, suggesting {\it ring}, is likely to have an analog
in Part~II with {\tt r} replaced by {\tt f}, suggesting {\it field}.

\section{Matrices}\label{mat}

The ACL2 events reported in this section are taken from the file {\tt rmat.lisp}, which begins with the definition of an {\tt m}$\times${\tt n} {\it matrix} {\tt a} over the ring {\tt R}
as a proper list of {\tt m} rlists, each of length {\tt n}:

\begin{small}
\begin{verbatim}
  (defun rmatp (a m n)
    (if (zp m)
        (null a)
      (and (consp a)
           (rlistnp (car a) n)
           (rmatp (cdr a) (1- m) n))))
\end{verbatim}
\end{small}
Each member of {\tt a} is a {\it row}; a {\it column} is constructed by extracting an entry from each row:

\begin{small}
\begin{verbatim}
  (defun row (i a) (nth i a))
  (defun col (j a)
    (if (consp a)
        (cons (nth j (car a)) (col j (cdr a)))
      ()))
\end{verbatim}
\end{small}
The entry of {\tt a}  in row {\tt i} and column {\tt j}:

\begin{small}
\begin{verbatim}
  (defun entry (i j a) (nth j (nth i a)))
\end{verbatim}
\end{small}
The basic operation of replacing row {\tt k} of {\tt a} with an rlist {\tt r}:

\begin{small}
\begin{verbatim}
 (defun replace-row (a k r)
   (if (zp k)
       (cons r (cdr a))
     (cons (car a) (replace-row (cdr a) (1- k) r))))
\end{verbatim}
\end{small}
If two {\tt m}$\times${\tt n} matrices are not equal, then some pair of corresponding entries are different.  The function {\tt entry-diff}
conducts a search and returns the row and column in which this occurs:

\begin{small}
\begin{verbatim}
  (defthmd rmat-entry-diff-lemma
    (implies (and (posp m) (posp n) (rmatp a m n) (rmatp b m n) (not (equal a b)))
             (let* ((pair (entry-diff a b)) (i (car pair)) (j (cdr pair)))
               (and (natp i) (< i m) (natp j) (< j n)
                    (not (equal (entry i j a) (entry i j b)))))))
\end{verbatim}
\end{small}
If we can prove that corresponding entries of {\tt a} and {\tt b} are equal, then we may invoke this result to conclude that
{\tt a} = {\tt b}.

The recursive definitions of the sum of two matrices, {\tt (rmat-add a b)}, and the product of a scalar and a matrix, {\tt (rmat-scalar-mul c a)}, are trival.
We shall also find it convenient to define the sum of the entries of a matrix in row-major order:

\begin{small}
\begin{verbatim}
  (defun rmat-sum (a)
    (if (consp a)
        (r+ (rlist-sum (car a)) (rmat-sum (cdr a)))
      (r0)))
\end{verbatim}
\end{small}
Matrix multiplication is a more complicated operation, deferred to Subsection~\ref{mul}.

\subsection{Transpose}

The {\it transpose} of a matrix is the list of its columns:

\begin{small}
\begin{verbatim}
  (defun transpose-mat-aux (a j n)
    (if (and (natp j) (natp n) (< j n))
        (cons (col j a) (transpose-mat-aux a (1+ j) n))
      ()))
  (defund transpose-mat (a) (transpose-mat-aux a 0 (len (car a))))
\end{verbatim}
\end{small}
We list some simple consequences of the definition:

\begin{small}
\begin{verbatim}
  (defthm transpose-rmat-entry
    (implies (and (posp m) (posp n) (rmatp a m n) (natp j) (< j n) (natp i) (< i m))
             (equal (entry j i (transpose-mat a))
                    (entry i j a))))
  (defthm transpose-rmat-2
    (implies (and (posp m) (posp n) (rmatp a m n))
             (equal (transpose-mat (transpose-mat a))
                    a)))
  (defthmd col-transpose-rmat
    (implies (and (posp m) (posp n) (rmatp a m n) (natp j) (< j m))
             (equal (col j (transpose-mat a))
                    (row j a))))
\end{verbatim}
\end{small}
The replacement of a column is now readily defined using the transpose:

\begin{small}
\begin{verbatim}
  (defund replace-col (a k r) (transpose-mat (replace-row (transpose-mat a) k r)))
\end{verbatim}
\end{small}

Our proof of associativity of matrix multiplication uses the observation that the entries of an {\tt m}$\times${\tt n} matrix {\tt a} have the same sum, as
computed by {\tt rmat-sum}, as those
of its transpose.  This is trivially true if either {\tt m} or {\tt n} is 0.  Otherwise, we derive the {\tt (m-1)x(n-1)} matrix {\tt (strip-mat a)} by deleting
the first row and the first column of {\tt a}, and prove the following:

\begin{small}
\begin{verbatim}
  (defthmd sum-rmat-strip-mat
    (implies (and (posp m) (posp n) (rmatp a m n))
             (equal (rmat-sum a)
                    (r+ (entry 0 0 a)
                        (r+ (r+ (rlist-sum (cdr (row 0 a)))
                                (rlist-sum (cdr (col 0 a))))
                            (rmat-sum (strip-mat a)))))))
\end{verbatim}
\end{small}
The desired lemma follows by induction, using {\tt sum-rmat-strip-mat} to rewrite both sides of the equation and {\tt col-transpose-rmat} to complete the proof:

\begin{small}
\begin{verbatim}
  (defthmd sum-rmat-transpose
    (implies (and (natp m) (natp n) (rmatp a m n))
             (equal (rmat-sum (transpose-mat a))
                    (rmat-sum a))))
\end{verbatim}
\end{small}

\subsection{Multiplication}\label{mul}

The product of matrices {\tt a} and {\tt b} is defined when the number of columns of {\tt a} is the number of rows of {\tt b}.
The product has the same number of rows as {\tt a} and the same number of columns as {\tt b}.  Each row of the product is the list of dot
products of the corresponding row of {\tt a} and the columns of {\tt b}:

\begin{small}
\begin{verbatim}
  (defund rmat* (a b)
    (if (consp a)
        (cons (rdot-list (car a) (transpose-mat b))
              (rmat* (cdr a) b))
      ()))
  (defthm rmatp-rmat*
    (implies (and (rmatp a m n) (rmatp b n p) (posp m) (posp n) (posp p))
             (rmatp (rmat* a b) m p)))
  (defthmd rmat*-entry
    (implies (and (posp m) (posp n) (posp p) (rmatp a m n) (rmatp b n p)
                  (natp i) (< i m) (natp j) (< j p))
             (equal (entry i j (rmat* a b))
                    (rdot (row i a) (col j b)))))
\end{verbatim}
\end{small}
The formula for the transpose of a product is an immediate consequence of {\tt transpose-rmat-entry}, {\tt rmat*-entry}, and
{\tt rmat-entry-diff-lemma}:

\begin{small}
\begin{verbatim}
  (defthmd transpose-rmat*
    (implies (and (posp m) (posp n) (posp p) (rmatp a m n) (rmatp b n p))
             (equal (transpose-mat (rmat* a b))
                    (rmat* (transpose-mat b) (transpose-mat a)))))
\end{verbatim}
\end{small}
For {\tt i} $<$ {\tt n}, row {\tt i} of the {\tt n}$\times${\tt n} {\it identity matrix} is the {\it unit vector} {\tt (runit i n)}, the rlist of
length {\tt n} with 1 at index {\tt i} and 0 elsewhere:

\begin{small}
\begin{verbatim}
  (defun runit (i n)
    (if (zp n) ()
      (if (zp i) (cons (r1) (rlistn0 (1- n)))
        (cons (r0) (runit (1- i) (1- n))))))
  (defun id-rmat-aux (i n)
    (if (and (natp i) (natp n) (< i n))
        (cons (runit i n) (id-rmat-aux (1+ i) n))
      ()))
  (defund id-rmat (n) (id-rmat-aux 0 n))
\end{verbatim}
\end{small}
The entries of the identity matrix are given by the {\it Kronecker delta} function:

\begin{small}
\begin{verbatim}
  (defun rdelta (i j) (if (= i j) (r1) (r0)))
  (defthmd entry-id-rmat
    (implies (and (natp n) (natp i) (natp j) (< i n) (< j n))
             (equal (entry i j (id-rmat n)) (rdelta i j))))
\end{verbatim}
\end{small}
It follows that the identity matrix is its own transpose, which in turn implies its defining properties:
\begin{small}
\begin{verbatim}
  (defthmd transpose-id-rmat
    (implies (natp n) (equal (transpose-mat (id-rmat n)) (id-rmat n))))
  (defthmd id-rmat-right
    (implies (and (posp m) (posp n) (rmatp a m n))
             (equal (rmat* a (id-rmat n)) a)))
  (defthmd id-rmat-left
    (implies (and (posp m) (posp n) (rmatp a m n))
             (equal (rmat* (id-rmat m) a) a)))
\end{verbatim}
\end{small}

To prove associativity of multiplication, let {\tt a}, {\tt b}, and {\tt c} be matrices of dimensions {\tt m}$\times${\tt n}, {\tt n}$\times${\tt p}, and
{\tt p}$\times${\tt q}, respectively, so that both products {\tt (rmat a (rmat* b c))} and {\tt (rmat* (rmat* a b) c)} are {\tt m}$\times${\tt q} matrices.  It will suffice
to show that corresponding entries agree:

\begin{small}
\begin{equation}\label{eqn0}
  \mbox{\tt (entry i j (rmat* a (rmat* b c))) = (entry i j (rmat* (rmat* a b) c))}.
\end{equation}
\end{small}
The usual informal proof proceeds by expanding the matrix products as well as the resulting dot products.  In standard notation (e.g., writing $a_{ir}$ for {\tt (entry i r a)}), the resulting goal is
\[
\sum_{r=0}^{n-1}\sum_{s=0}^{p-1}a_{ir}b_{rs}c_{sj} = \sum_{s=0}^{p-1}\sum_{r=0}^{n-1}a_{ir}b_{rs}c_{sj}.
\]
The proof is completed by simply observing that the sum on the right is a rearrangement of the three-way products that appear in the sum on the left.  Our objective is a formal proof that captures
the intuition underlying this observation.

We shall show that these products are the entries of the {\tt n}$\times${\tt p} matrix {\tt (rmat12 a b c i j)}, defined as follows:

\begin{small}
\begin{verbatim}
  (defun rlist-mul-list (x l)
    (if (consp l)
        (cons (rlist-mul x (car l))
              (rlist-mul-list x (cdr l)))
      ()))
  (defun rlist-scalar-mul-list (x l)
    (if (consp l)
        (cons (rlist-scalar-mul (car x) (car l))
              (rlist-scalar-mul-list (cdr x) (cdr l)))
      ()))
  (defund rmat12 (a b c i j)
    (rlist-scalar-mul-list (row i a) (rlist-mul-list (col j c) b)))
\end{verbatim}
\end{small}
To compute the entries of this matrix, first we compute its {\tt r}th row:

\begin{small}
\begin{verbatim}
  (nth r (rmat12 a b c i j))
    = (rlist-scalar-mul (nth r (row i a)) (nth r (rlist-mul-list (col j c) b)))
    = (rlist-scalar-mul (entry i r a) (rlist-mul (col j c) (nth r b)))
\end{verbatim}
\end{small}
Now the {\tt s}th entry of the {\tt r}th row:
\begin{small}
\begin{verbatim}
  (entry r s (rmat12 a b c i j))
    = (nth s (nth r (rmat12 a b c i j)))
    = (nth s (rlist-scalar-mul (entry i r a) (rlist-mul (col j c) (nth r b))))
    = (entry i r a) * (nth s (rlist-mul (col j c) (nth r b)))
    = (entry i r a) * ((nth s (col j c)) * (nth s (nth r b)))
    = (entry i r a) * ((entry s j c) * (entry r s b))
    = (entry i r a) * (entry r s b) * (entry s j c)
\end{verbatim}
\end{small}
Next we compute  {\tt (rmat-sum (rmat12 a b c i j))}.  As a first step, it is easily 
shown by induction that if {\tt x} is an rlist of length {\tt n} and {\tt l} is a matrix with {\tt n} rows, then

\begin{small}
\begin{verbatim}
  (rmat-sum (rlist-scalar-mul-list x l)) = (rdot x (rlist-sum-list l)).
\end{verbatim}
\end{small}
We apply this result to the definition of {\tt rmat-sum}, substituting {\tt (row i a)} for {\tt x} and {\tt (rlist-mul-list (col j c) b)} for {\tt l}.  This yields the following expression for
{\tt rmat-sum (rmat12 a b c i j))}:
\begin{small}
\begin{verbatim}
  (rdot (row i a) (rlist-sum-list (rlist-mul-list (col j c) b))).
\end{verbatim}
\end{small}
Note that {\tt (rlist-sum-list (rlist-mul-list (col j c) b))} and {\tt (col j (rmat* b c))} are both rlists of length {\tt n}.
To prove equality, it suffices to show that corresponding members are equal:

\begin{small}
\begin{verbatim}
  (nth k (rlist-sum-list (rlist-mul-list (col j c) b)))
    = (rlist-sum (nth k (rlist-mul-list (col j c) b)))
    = (rlist-sum (rlist-mul (col j c) (nth k b)))
    = (rdot (col j c) (nth k b))
    = (rdot (col j c) (row k b))
    = (rdot (row k b) (col j c))
    = (entry k j (rmat* b c))
    = (nth k (col j (rmat* b c)))
\end{verbatim}
\end{small}
Thus, {\tt (rlist-sum-list (rlist-mul-list (col j c) b))} = {\tt (col j (rmat* b c))}.  It follows that
\begin{small}
\begin{verbatim}
  (rmat-sum (rmat12 a b c i j)) = (rdot (row i a) (col j (rmat* b c)))
                                = (entry i j (rmat* a (rmat* b c))):
\end{verbatim}
\end{small}

The {\tt p}$\times${\tt n} matrix corresponding to the right side of Equation~(\ref{eqn0}) is similarly defined:

\begin{small}
\begin{verbatim}
  (defund rmat21 (a b c i j)
    (rlist-scalar-mul-list (col j c)
                           (rlist-mul-list (row i a) (transpose-mat b))))
\end{verbatim}
\end{small}
Minor variations in the above derivations yield an expression for the entries of this matrix,
\begin{small}
\begin{verbatim}
  (entry r s (rmat21 a b c i j)) = (r* (entry i s a) (r* (entry s r b) (entry r j c))),
\end{verbatim}
\end{small}
and the sum of these entries:

\begin{small}
\begin{verbatim}
  (rmat-sum (rmat21 a b c i j)) = (entry i j (rmat* (rmat* a b) c)).
\end{verbatim}
\end{small}
Thus, {\tt (entry r s (rmat21 a b c i j))} = {\tt  (entry s r (rmat12 a b c i j))}, and hence

\begin{small}
\begin{verbatim}
  (transpose-mat (rmat12 a b c i j)) = (rmat21 a b c i j).
\end{verbatim}
\end{small}
Finally, Equation~(\ref{eqn0}) follows from {\tt sum-rmat-transpose}, and associativity holds:

\begin{small}
\begin{verbatim}
  (defthmd rmat*-assoc
    (implies (and (rmatp a m n) (rmatp b n p) (rmatp c p q)
                  (posp m) (posp n) (posp p) (posp q))
             (equal (rmat* a (rmat* b c))
                    (rmat* (rmat* a b) c))))
\end{verbatim}
\end{small}

\section{Determinants}\label{det}

In {\tt rdet.lisp}, we formalize the classical definition of the {\it determinant} of an {\tt n}$\times${\tt n} matrix over the ring {\tt R}, based on the symmetric group
{\tt (sym n)} as defined in {\tt books/projects/groups/symmetric.lisp} and documented in \cite{groups3}.  The elements of this group are the members of the list {\tt (slist n)} of permutations
of the list {\tt (ninit n)} = {\tt (0 1 ... n-1)}.  Such a permutation {\tt p} may be viewed as a bijection of {\tt (ninit n)} that maps an index {\tt j} to
{\tt (nth j p)}.  The composition of permutations {\tt p} and {\tt q} is computed by the group operation, {\tt (comp-perm p q n)}.  Note that {\tt (ninit n)}
itself is the group identity.

A {\it transposition} is a permution, denoted by {\tt (transpose i j n)}, that simply interchanges two distinct indices {\tt i} and {\tt j}.  Every permutation
may be represented as a composition of a list of transpositions, and while neither this list nor its length is unique, its length is either always even or always odd
for a given permutation {\tt p}; {\tt p} is said to be {\it even} or {\it odd} accordingly.

A permutation {\tt p} is applied to an arbitrary list {\tt l} of length {\tt n} by the following function:

\begin{small}
\begin{verbatim}
  (defun permute (l p)
    (if (consp p)
        (cons (nth (car p) l) (permute l (cdr p)))
      ()))
\end{verbatim}
\end{small}
A critical property of {\tt permute} pertains to a product of permutations:

\begin{small}
\begin{verbatim}
  (defthm permute-comp-perm
    (implies (and (true-listp l) (consp l) (in x (sym (len l))) (in y (sym (len l))))
             (equal (permute (permute l x) y)
                    (permute l (comp-perm x y (len l))))))
\end{verbatim}
\end{small}

Each permutation {\tt p} in {\tt (sym n)} contributes a term {\tt (rdet-term a p n)} to the determinant of an {\tt n}$\times${\tt n} matrix {\tt a}, computed as follows:

\begin{itemize}
\item [(1)] For each {\tt i} $<$ {\tt n}, select the entry of {\tt (row i a)} in column {\tt (nth i p)};
\item [(2)] Compute the product of these {\tt n} entries;
\item [(3)] Negate the product if {\tt p} is an odd permutation.
\end{itemize}

\begin{small}
\begin{verbatim}
  (defun rdet-prod (a p n)
    (if (zp n)
        (r1)
      (r* (rdet-prod a p (1- n))
          (entry (1- n) (nth (1- n) p) a))))
  (defund rdet-term (a p n)
    (if (even-perm-p p n)
        (rdet-prod a p n)
      (r- (rdet-prod a p n))))
\end{verbatim}
\end{small}
The determinant of {\tt a} is the the sum over {\tt (slist n)} of these signed products:

\begin{small}
\begin{verbatim}
  (defun rdet-sum (a l n)
    (if (consp l)
        (r+ (rdet-term a (car l) n) (rdet-sum a (cdr l) n))
      (r0)))
  (defund rdet (a n) (rdet-sum a (slist n) n))
\end{verbatim}
\end{small}

\subsection{Properties}

To compute the determinant of the identity matrix, note that if {\tt p} is any permutation other than the identity {\tt (ninit n)}, we can find {\tt i} $<$ {\tt n}
such that {\tt (nth i p)} $\neq$ {\tt i}, and hence {\tt (entry i (nth i p) (id-rmat n))} = 0, which implies {\tt (rdet-term (id-rmat n) p n)} = 0.  On the
other hand, {\tt (nth i (ninit n))} = {\tt i} for all {\tt i}, which implies {\tt (rdet-term (id-rmat n) (ninit n) n)} = 1.  Thus,

\begin{small}
\begin{verbatim}
  (defthm rdet-id-rmat (implies (posp n) (equal (rdet (id-rmat n) n) (r1))))
\end{verbatim}
\end{small}
The determinant is invariant under {\tt transpose-mat}.  This follows from the observation that the term contributed to the determinant of the transpose of
{\tt a} by a permutation {\tt p} is the same as the term contributed to the determinant of {\tt a} by the inverse of {\tt p}:

\begin{small}
\begin{verbatim}
  (defthmd rdet-transpose
    (implies (and (posp n) (rmatp a n n))
             (equal (rdet (transpose-mat a) n) (rdet a n))))
\end{verbatim}
\end{small}
If every entry of the {\tt k}th row of {\tt a} is 0, then for all {\tt p}, the {\tt k}th factor of {\tt (rdet-prod a p n)} is 0,
and it follows that the determinant of {\tt a} is 0:

\begin{small}
\begin{verbatim}
  (defthmd rdet-row-0
    (implies (and (rmatp a n n) (posp n) (natp k) (< k n) (= (nth k a) (rlistn0 n)))
             (equal (rdet a n) (r0))))
\end{verbatim}
\end{small}
Furthermore, the determinant is {\it alternating}, i.e., if two rows of {\tt a} are equal, then its determinant is 0.  To prove
this, suppose rows {\tt i} and {\tt j} are equal, where {\tt i} $\neq$ {\tt j}.  Given a permutation {\tt p}, let
{\tt p'} = {\tt (comp-perm p (transpose i j n) n)}.  The factors of {\tt (rdet-prod a p' n)} are the same as those of
{\tt (rdet-prod a p n)}.  But {\tt p} and {\tt p'} have opposite parities, and therefore {\tt (rdet-term a p' n)} is the negative
of {\tt (rdet-term a p n)}.  Consequently, the sum of terms contributed by the odd permutations is the negative of the sum of terms
contributed by the even permutations, and we have

\begin{small}
\begin{verbatim}
  (defthmd rdet-alternating
    (implies (and (rmatp a n n) (posp n) 
                  (natp i) (< i n) (natp j) (< j n) (not (= i j))
                  (= (row i a) (row j a)))
             (equal (rdet a n) (r0))))
\end{verbatim}
\end{small}
The determinant is also {\it n-linear}, i.e., linear as a function of each row.  This property is specified in terms of the {\tt replace-row} operation.
For a given row {\tt i} and permutation {\tt p}, the term contributed by {\tt p} to the determinant of {\tt (replace-row a i x)} is a linear function of {\tt x}:

\begin{small}
\begin{verbatim}
  (defthm rdet-term-replace-row
    (implies (and (rmatp a n n) (posp n) (member p (slist n))
                  (rlistnp x n) (rlistnp y n) (rp c)
                  (natp i) (< i n))
             (let ((a1 (replace-row a i x))
                   (a2 (replace-row a i y))
                   (a3 (replace-row a i (rlist-add (rlist-scalar-mul c x) y))))
               (equal (rdet-term a3 p n)
                      (r+ (r* c (rdet-term a1 p n)) (rdet-term a2 p n))))))
\end{verbatim}
\end{small}
The desired result follows by summing over all permutations:

\begin{small}
\begin{verbatim}
  (defthm rdet-n-linear
    (implies (and (rmatp a n n) (posp n) (natp i) (< i n)
                  (rlistnp x n) (rlistnp y n) (rp c))
             (equal (rdet (replace-row a i (rlist-add (rlist-scalar-mul c x) y)) n)
                    (r+ (r* c (rdet (replace-row a i x) n))
                        (rdet (replace-row a i y) n)))))
\end{verbatim}
\end{small}

\subsection{Uniqueness}

We shall show that {\tt rdet} is the unique {\tt n}-linear alternating function on {\tt n}$\times${\tt n} matrices that satisfies {\tt (rdet (id-rmat n) n)} = 1.
To that end, we introduce a constrained function {\tt rdet0} as follows:

\begin{small}
\begin{verbatim}
  (encapsulate (((rdet0 * *) => *))
    (local (defun rdet0 (a n) (rdet a n)))
    (defthm rp-rdet0
      (implies (and (rmatp a n n) (posp n))
               (rp (rdet0 a n))))
    (defthmd rdet0-n-linear
      (implies (and (rmatp a n n) (posp n) (natp i) (< i n)
                    (rlistnp x n) (rlistnp y n) (rp c))
               (equal (rdet0 (replace-row a i (rlist-add (rlist-scalar-mul c x) y)) n)
                      (r+ (r* c (rdet0 (replace-row a i x) n))
                          (rdet0 (replace-row a i y) n)))))
    (defthmd rdet0-adjacent-equal
      (implies (and (rmatp a n n) (posp n)
                    (natp i) (< i (1- n)) (= (row i a) (row (1+ i) a)))
               (equal (rdet0 a n) (r0)))))
\end{verbatim}
\end{small}
Our main objective is to prove that

\begin{small}
\begin{equation}\label{eqn1}
\mbox{\tt (rdet0 a n) = (r* (rdet a n) (rdet0 (id-rmat n)))}.
\end{equation}
\end{small}
If we then prove that a given function {\tt (f a n)} satisfies the constraints on {\tt rdet0}, then we
may conclude by functional instantiation that {\tt (f a n)} = {\tt (r* (rdet a n) (f (id-rmat n) n))}.  From this it will follow that if
{\tt f} has the additional property {\tt (f (id-rmat n) n)} = 1, then {\tt (f a n)} = {\tt (rdet a n)}.

Note that instead of assuming that {\tt rdet0} is alternating, we have imposed the weaker constraint {\tt rdet0-adjacent-equal}, which
says that the value is 0 if two {\it adjacent} rows are equal.  This relaxes the proof obligations for functional instantiation, which
will be critical for the proof of correctness of cofactor expansion (Section~\ref{cofactors}).  However, it is a consequence of the above
constraints that {\tt rdet0} is alternating.  To establish this, we first show by a sequence of applications of {\tt rdet0-n-linear} and
{\tt rdet0-adjacent-equal} that transposing two adjacent rows negates the value of {\tt rdet0}.  It is also easily shown that an arbitrary
transposition may be expressed as a composition of an odd number of transpositions of adjacent rows, and it follows that the value is
negated by transposing any two rows:

\begin{small}
\begin{verbatim}
  (defthmd rdet0-permute-transpose
    (implies (and (rmatp a n n) (posp n)
                  (natp i) (natp j) (< i j) (< j n))
             (equal (rdet0 (permute a (transpose i j n)) n)
                    (r- (rdet0 a n)))))
\end{verbatim}
\end{small}
Since every permutation is a product of transpositions, this yields the following generalization:

\begin{small}
\begin{verbatim}
  (defthmd rdet0-permute-rows
    (implies (and (rmatp a n n) (posp n) (in p (sym n)))
             (equal (rdet0 (permute a p) n)
                    (if (even-perm-p p n)
                        (rdet0 a n)
                      (r- (rdet0 a n))))))
\end{verbatim}
\end{small}

Now suppose {\tt (row i a)} = {\tt (row j a)}, where 0 $\leq$ {\tt i} $<$ {\tt j} $<$ {\tt n}.  By {\tt rdet0-adjacent-equal}, we may also assume
{\tt i} + 1 $<$ {\tt j}.  Let {\tt a'} = {\tt (permute (transpose (1+ i) j n) a)}.  Then

\begin{small}
\begin{verbatim}
  (nth (1+ i) a') = (nth j a) = (nth i a) = (nth i a').
\end{verbatim}
\end{small}
By {\tt rdet0-adjacent-equal}, {\tt (rdet0 a')} = 0, and by {\tt rdet0-permute-transpose},

\begin{small}
\begin{verbatim}
  (rdet0 a n) = (r- (rdet0 a' n) = (r- 0) = 0.
\end{verbatim}
\end{small}
Thus, {\tt rdet0} is an alternating function:

\begin{small}
\begin{verbatim}
  (defthmd rdet0-alternating
    (implies (and (rmatp a n n) (posp n) (natp i) (natp j) (< i n) (< j n)
                  (not (= i j)) (= (row i a) (row j a)))
             (equal (rdet0 a n) (r0))))
\end{verbatim}
\end{small}

Our proof of Equation~(\ref{eqn1}) involves arbitrary lists of length {\tt k} $\leq$ {\tt n} of natural numbers less than {\tt n}, which we call {\tt k}-{\it tuples}.
We begin with the following definitions:

\begin{itemize}
\item {\tt (tuplep x k n)} is a predicate that recognizes a {\tt k}-tuple;
\item {\tt (extend-tuple x n)} returns the list of {\tt n} ({\tt k}+1)-tuples constructed from a given {\tt k}-tuple {\tt x} by appending each
natural number less than {\tt n};
\item {\tt (extend-tuples l n)} returns the list of all ({\tt k}+1)-tuples constructed in this way from the members of a list {\tt l} of {\tt k}-tuples.
\end{itemize}
The list of all {\tt k}-tuples is defined recursively:

\begin{small}
\begin{verbatim}
(defun all-tuples (k n)
  (if (zp k)
      (list ())
    (extend-tuples (all-tuples (1- k) n) n)))
\end{verbatim}
\end{small}
Let {\tt a} be a fixed {\tt n}$\times${\tt n} matrix.  We associate a value {\tt (reval-tuple x k a n)} with each {\tt k}-tuple {\tt x} as follows.  First we construct an rlist of length {\tt k}, {\tt (extract-entries x a)}, the {\tt j}th member of which is {\tt (entry j (nth j x) a)}:

\begin{small}
\begin{verbatim}
  (defun extract-entries (x a)
    (if (consp x)
        (cons (nth (car x) (car a))
              (extract-entries (cdr x) (cdr a)))
      ()))
\end{verbatim}
\end{small}
We define {\tt (runits x n)} to be the list of unit vectors corresponding to the members of {\tt x}:

\begin{small}
\begin{verbatim}
  (defun runits (x n)
    (if (consp x)
        (cons (runit (car x) n) (runits (cdr x) n))
      ()))
\end{verbatim}
\end{small}
The value {\tt (reval-tuple x k a n)} is the product of the members of {\tt (extract-entries x a)} together with the value of {\tt rdet0}
applied to the matrix derived from {\tt a} by replacing its first {\tt k} rows with {\tt (runits x n)}:

\begin{small}
\begin{verbatim}
  (defun reval-tuple (x k a n)
    (r* (rlist-prod (extract-entries x a))
        (rdet0 (append (runits x n) (nthcdr k a)) n)))
\end{verbatim}
\end{small}
We also define the sum of the values of {\tt (reval-tuple x k a n)} as {\tt x} ranges over a list {\tt l} of {\tt k}-tuples:

\begin{small}
\begin{verbatim}
  (defun rsum-tuples (l k a n)
    (if (consp l)
        (r+ (reval-tuple (car l) k a n) (rsum-tuples (cdr l) k a n))
      (r0)))
\end{verbatim}
\end{small}
We would like to compute {\tt (rsum-tuples (all-tuples k n) k a n)}.  Since the only member of {\tt (all-tuples 0 n)} is {\tt NIL}, the case {\tt k} = 0 is trivial:

\begin{small}
\begin{equation}\label{eqn2}
  \mbox{\tt (rsum-tuples (all-tuples 0 n) 0 a n) = (reval-tuple () 0 a n) = (rdet0 a n)}.
\end{equation}
\end{small}
For the case {\tt k} = {\tt n}, we observe that {\tt (nthcdr n a)} = {\tt NIL} and that if the members of {\tt x} are not distinct, then the matrix {\tt (runits x n)}
has two equal rows and by {\tt rdet0-alternating}, {\tt (rdet0 (runits x n) n)} = 0.  Thus, in the computation of {\tt (rsum-tuples (all-tuples n n) n a n)}, we need only
consider the {\tt n}-tuples that are permutations.  If {\tt p} is in {\tt (sym n)}, then by {\tt rdet0-permute-rows},

\begin{small}
\begin{verbatim}
  (rdet0 (runits p n) n)
    = (rdet0 (permute (id-rmat n) p) n)
    = (if (even-perm-p n n) (rdet0 (id-rmat n) n) (r- (rdet0 (id-rmat n) n)))
\end{verbatim}
\end{small}
and {\tt (extract-entries p a)} = {\tt (rdet-prod a p n)}.  Consequently,

\begin{small}
\begin{verbatim}
  (reval-tuple p n a n) = (r* (rdet-term a p n) (rdet0 (id-rmat n) n)).
\end{verbatim}
\end{small}
Summing over {\tt (slist n)}, we have
\begin{small}
\begin{equation}\label{eqn3}
  \mbox{\tt (rsum-tuples (all-tuples n n) n a n) = (r* (rdet a n) (rdet0 (id-rmat n) n))}.
\end{equation}
\end{small}
For 0 $\leq$ {\tt k} $<$ {\tt n} and {\tt (tuplep x k n)}, repeated application of {\tt rdet0-n-linear} yields

\begin{small}
\begin{verbatim}
  (rsum-tuples (extend-tuple x) (1+ k) a n) = (reval-tuple x k a n).
\end{verbatim}
\end{small}
Summing over {\tt (all-tuples k n)}, we have the recurrence formula

\begin{small}
\begin{verbatim}
(rsum-tuples (all-tuples (1+ k) n) (1+ k) a n) = (rsum-tuples (all-tuples k n) k a n).
\end{verbatim}
\end{small}
By induction, {\tt (rsum-tuples (all-tuples k n) k a n)} is independent of {\tt k}.  In particular,

\begin{small}
\begin{verbatim}
  (rsum-tuples (all-tuples n n) n a n) = (rsum-tuples (all-tuples 0 n) 0 a n). 
\end{verbatim}
\end{small}
Equation~(\ref{eqn1}) follows from this result together with Equations (\ref{eqn2}) and (\ref{eqn3}):

\begin{small}
\begin{verbatim}
  (defthmd rdet-unique
    (implies (rmatp a n n)
             (equal (rdet0 a n)
                    (r* (rdet a n) (rdet0 (id-rmat n) n)))))
\end{verbatim}
\end{small}

\subsection{Multiplicativity}

If we had further constrained the function {\tt rdet0} to satisfy {\tt (rdet0 (id-rmat n) n)} = 1, then we could have replaced the conclusion of {\tt rdet-unique}
with the simpler equation {\tt (rdet0 a n)} = {\tt (rdet a n)}.  One reason behind our weaker specification is that it
allows us to prove the multiplicativity property, {\tt (rdet (rmat* a b) n) = (r* (rdet a n) (rdet b n)}, by functional instantiation.  We define

\begin{small}
\begin{verbatim}
  (defun rdet-rmat* (a b n) (rdet (rmat* a b) n))
\end{verbatim}
\end{small}
Our goal is the functional instance of {\tt rdet-unique} derived by substituting

\begin{small}
\begin{verbatim}
  (lambda (a n) (rdet-rmat* a b n))
\end{verbatim}
\end{small}
for {\tt rdet0}.  This requires that we prove the analogs of the two nontrivial constraints on {\tt rdet0}.  The first is a consequence of {\tt rdet-n-linear} and the definitions
of {\tt rmat*}, {\tt rdot-list}, and {\tt rlist-scalar-mul}:

\begin{small}
\begin{verbatim}
  (defthmd rdet-rmat*-n-linear
    (implies (and (rmatp a n n) (rmatp b n n) (posp n) (natp k) (< k n)
                  (rlistnp x n) (rlistnp y n) (rp c))
             (equal (rdet-rmat* (replace-row a k (rlist-add (rlist-scalar-mul c x) y))
                                b n)
                    (r+ (r* c (rdet-rmat* (replace-row a k x) b n))
                        (rdet-rmat* (replace-row a k y) b n)))))
\end{verbatim}
\end{small}
The second follows from {\tt rdet-alternating} and the observation that if {\tt (row k a)} = {\tt (row (1+ k) a)}, then {\tt (row k (rmat* a b))} = {\tt (row (1+ k) (rmat* a b))}:

\begin{small}
\begin{verbatim}
  (defthmd rdet-rmat*-adjacent-equal
    (implies (and (rmatp a n n) (rmatp b n n) (posp n)
                  (natp k) (< k (1- n)) (= (row k a) (row (1+ k) a)))
             (equal (rdet-rmat* a b n) (r0))))
\end{verbatim}
\end{small}
Functional instantiation of {\tt rdet-unique} yields

\begin{small}
\begin{verbatim}
  (rdet-rmat* a b n) = (r* (rdet a n) (rdet-rmat* (id-rmat n) b n)).
\end{verbatim}
\end{small}
Expanding {\tt rdet-rmat*} and applying {\tt id-rmat-left}, we have

\begin{small}
\begin{verbatim}
  (defthmd rdet-multiplicative
    (implies (and (rmatp a n n) (rmatp b n n) (posp n))
             (equal (rdet (rmat* a b) n)
                    (r* (rdet a n) (rdet b n)))))
\end{verbatim}
\end{small}

\section{Cofactors}\label{cofactors}

Given an {\tt n}$\times${\tt n} matrix {\tt a}, we define the {\tt (n-1)}$\times${\tt (n-1)} submatrix {\tt (minor i j a)} to be the result of deleting the {\tt i}th row and the {\tt j}th
column of {\tt a}:

\begin{small}
\begin{verbatim}
  (defun delete-row (k a)
    (if (zp k) (cdr a)
      (cons (car a) (delete-row (1- k) (cdr a)))))
  (defund delete-col (k a) (transpose-mat (delete-row k (transpose-mat a))))
  (defund minor (i j a) (delete-col j (delete-row i a)))
\end{verbatim}
\end{small}
Its entries may be computed as follows:

\begin{small}
\begin{verbatim}
  (defthmd entry-rmat-minor
    (implies (and (rmatp a n n) (natp n) (> n 1) (natp i) (natp j) (< i n) (< j n)
                  (natp r) (natp s) (< r (1- n)) (< s (1- n)))
           (equal (entry r s (minor i j a))
                  (entry (if (>= r i) (1+ r) r) (if (>= s j) (1+ s) s) a))))
\end{verbatim}
\end{small}
The {\it cofactor} of an entry of {\tt a} is the determinant of its minor with an attached sign determined by the parity of the sum of its indices:

\begin{small}
\begin{verbatim}
  (defund rdet-cofactor (i j a n)
    (if (evenp (+ i j))
        (rdet (minor i j a) (1- n))
      (r- (rdet (minor i j a) (1- n)))))
\end{verbatim}
\end{small}

\subsection{Cofactor Expansion}

The cofactor expansion of the determinant of {\tt a} by a column is computed by multiplying each entry of the column by its cofactor and summing the products:

\begin{small}
\begin{verbatim}
  (defun expand-rdet-col-aux (a i j n)
    (if (zp i) (r0)
      (r+ (r* (entry (1- i) j a) (rdet-cofactor (1- i) j a n))
          (expand-rdet-col-aux a (1- i) j n))))
  (defund expand-rdet-col (a j n) (expand-rdet-col-aux a n j n))
\end{verbatim}
\end{small}
Cofactor expansion by a row is similarly defined:

\begin{small}
\begin{verbatim}
  (defun expand-rdet-row-aux (a i j n)
    (if (zp j) (r0)
      (r+ (r* (entry i (1- j) a) (rdet-cofactor i (1- j) a n))
          (expand-rdet-row-aux a i (1- j) n))))
  (defund expand-rdet-row (a i n) (expand-rdet-row-aux a i n n))
\end{verbatim}
\end{small}
It follows from {\tt entry-rmat-minor} and {\tt transpose-rmat-entry} that

\begin{small}
\begin{verbatim}
  (transpose-mat (minor i j a)) = (minor j i (transpose-mat a)),
\end{verbatim}
\end{small}
which, in combination with {\tt rdet-transpose}, implies

\begin{small}
\begin{verbatim}
  (rdet-cofactor j i (transpose-mat a) n) = (rdet-cofactor i j a n).
\end{verbatim}
\end{small}
Consequently, cofactor expansion by column {\tt i} is equivalent to expansion of the transpose by  row {\tt i}:

\begin{small}
\begin{verbatim}
  (defthmd expand-rdet-row-transpose
    (implies (and (rmatp a n n) (natp n) (> n 1) (natp i) (< i n))
             (equal (expand-rdet-row (transpose-mat a) i n)
                    (expand-rdet-col a i n))))
\end{verbatim}
\end{small}
We shall prove, by functional instantiation of {\tt rdet-unique}, that the result of cofactor expansion by a column has the same value as the determinant,
and it will follow that the same is true for expansion by a row.  Once again, this requires proving analogs of the constraints on {\tt rdet0}.

It is clear that replacing row {\tt i} of {\tt a} does not alter {\tt (rdet-cofactor i j a b)}.
On the other hand, for {\tt k} $\neq$ {\tt i}, {\tt (rdet-cofactor i j a n)} is a linear function of {\tt (row k a)}:

\begin{small}
\begin{verbatim}
  (defthmd rdet-cofactor-n-linear
    (implies (and (rmatp a n n) (natp n) (> n 1) (natp i) (< i n) (natp j) (< j n)
                  (natp k) (< k n) (not (= k i)) (rlistnp x n) (rlistnp y n) (rp c))
             (equal (rdet-cofactor
                       i j (replace-row a k (rlist-add (rlist-scalar-mul c x) y)) n)
                    (r+ (r* c (rdet-cofactor i j (replace-row a k x) n))
                        (rdet-cofactor i j (replace-row a k y) n)))))
\end{verbatim}
\end{small}
It follows that cofactor expansion by column {\tt j} is {\tt n}-linear:

\begin{small}
\begin{verbatim}
  (defthmd expand-rdet-col-n-linear
    (implies (and (rmatp a n n) (natp n) (> n 1) (natp j) (< j n)
                  (natp k) (< k n) (rlistnp x n) (rlistnp y n) (rp c))
             (equal (expand-rdet-col
                      (replace-row a k (rlist-add (rlist-scalar-mul c x) y)) j n)
                    (r+ (r* c (expand-rdet-col (replace-row a k x) j n))
                        (expand-rdet-col (replace-row a k y) j n)))))
\end{verbatim}
\end{small}

Now suppose adjacent rows {\tt k} and {\tt k} + 1 are equal.  Then for any index {\tt i} other than {\tt k} or {\tt k} + 1, {\tt (minor i j a)} has two
equal adjacent rows, and therefore {\tt (rdet-cofactor i j a n)} = 0.  Meanwhile,

\begin{small}
\begin{verbatim}
  (minor k j) = (minor (1+ k) j)
\end{verbatim}
\end{small}
and

\begin{small}
\begin{verbatim}
  (entry k j a) = (entry (1+ k) j a),
\end{verbatim}
\end{small}
but {\tt k} + {\tt j} and ({\tt k} + 1) + {\tt j} have opposite parities, and hence

\begin{small}
\begin{verbatim}
  (rdet-cofactor k j a n) + (rdet-cofactor (1+ k) j a n) = 0.
\end{verbatim}
\end{small}
Therefore, {\tt (expand-rdet-col a j n)} = 0:

\begin{small}
\begin{verbatim}
  (defthmd expand-rdet-col-adjacent-equal
    (implies (and (rmatp a n n) (> n 1) (natp j) (< j n)
                  (natp k) (< k (1- n)) (= (row k a) (row (1+ k) a)))
             (equal (expand-rdet-col a j n) (r0))))
\end{verbatim}
\end{small}
Thus, the constraints on {\tt rdet0} are satisfied, and by functional instantiation of {\tt rdet-unique}, we have the following:

\begin{small}
\begin{verbatim}
  (defthmd expand-rdet-col-val
    (implies (and (rmatp a n n) (posp n) (> n 1) (natp k) (< k n))
             (equal (expand-rdet-col a k n)
                    (r* (rdet a n) (expand-rdet-col (id-rmat n) k n)))))
\end{verbatim}
\end{small}

It remains to show that {\tt (expand-rdet-col (id-rmat n) k n)} = 1.  By {\tt row-rmat-minor} (see {\tt rdet.lisp}), we have the following expression for a row of
{\tt (minor i j (id-rmat n))}:

\begin{small}
\begin{verbatim}
  (defthmd nth-minor-id-rmat
    (implies (and (natp n) (> n 1) (natp i) (< i n) (natp j) (< j n)
                  (natp r) (< r (1- n)))
             (equal (nth r (minor i j (id-rmat n)))
                    (delete-nth j (runit (if (< r i) r (1+ r)) n)))))
\end{verbatim}
\end{small}
The following is a consequence of the definitions of {\tt runit} and {\tt delete-nth}:

\begin{small}
\begin{verbatim}
  (defthmd delete-nth-runit
    (implies (and (posp n) (natp j) (< j n) (natp k) (< k n))
             (equal (delete-nth j (runit k n))
                    (if (< j k) (runit (1- k) (1- n))
                      (if (> j k) (runit k (1- n))
                        (rlistn0 (1- n)))))))
\end{verbatim}
\end{small}
Consequently, if {\tt i} $\neq$ {\tt j}, then we find a zero row of {\tt (minor i j (id-rmat n))}, and by {\tt rdet-row-0}, its
determinant is 0.  On the other hand, {\tt (minor j j (id-rmat n))} = {\tt (id-rmat (1- n))} and the corresponding cofactor
is 1, as is the cofactor expansion:

\begin{small}
\begin{verbatim}
  (defthmd expand-rdet-col-id-rmat
    (implies (and (rmatp a n n) (natp n) (> n 1) (natp j) (< j n))
             (equal (expand-rdet-col (id-rmat n) j n) (r1))))
\end{verbatim}
\end{small}
Combining this with {\tt expand-rdet-col-val}, we have the correctness theorem for column expansion:

\begin{small}
\begin{verbatim}
  (defthmd expand-rdet-col-rdet
    (implies (and (rmatp a n n) (posp n) (> n 1) (natp k) (< k n))
             (equal (expand-rdet-col a k n) (rdet a n))))
\end{verbatim}
\end{small}
It follows from {\tt rdet-transpose}, {\tt expand-rdet-row-transpose}, and {\tt transpose-rmat-2} that the same holds for row expansion:

\begin{small}
\begin{verbatim}
  (defthmd expand-rdet-row-rdet
    (implies (and (rmatp a n n) (posp n) (> n 1) (natp k) (< k n))
             (equal (expand-rdet-row a k n) (rdet a n))))
\end{verbatim}
\end{small}

As a consequence of {\tt expand-rdet-row-rdet}, we have a recursive version of {\tt rdet}, based on cofactor expansion with respect to row 0:

\begin{small}
\begin{verbatim}
  (mutual-recursion
    (defund rdet-rec-cofactor (j a n)
      (if (zp n) ()
        (if (evenp j) (rdet-rec (minor 0 j a) (1- n))
          (r- (rdet-rec (minor 0 j a) (1- n))))))
    (defun expand-rdet-rec-aux (a j n)
      (if (zp j) (r0)
        (r+ (r* (entry 0 (1- j) a) (rdet-rec-cofactor (1- j) a n))
            (expand-rdet-rec-aux a (1- j) n))))
    (defund expand-rdet-rec (a n) (expand-rdet-rec-aux a n n))
    (defun rdet-rec (a n)
      (if (zp n) (r0)
        (if (= n 1) (entry 0 0 a)
          (expand-rdet-rec a n)))))
\end{verbatim}
\end{small}
The equivalence follows from {\tt expand-rdet-row-rdet} by induction (see {\tt rdet.lisp} for details):

\begin{small}
\begin{verbatim}
  (defthmd rdet-rec-rdet
    (implies (and (rmatp a n n) (posp n))
             (equal (rdet-rec a n) (rdet a n))))
\end{verbatim}
\end{small}

\subsection{Classical Adjoint}

We shall define the {\it cofactor matrix} of an {\tt n}$\times${\tt n} matrix {\tt a} to be the {\tt n}$\times${\tt n} matrix with entries

\begin{small}
\begin{verbatim}
  (entry i j (cofactor-rmat a b)) = (rdet-cofactor i j a n).
\end{verbatim}
\end{small}
To define this matrix, we first define a function that computes its {\tt i}th row:

\begin{small}
\begin{verbatim}
  (defun cofactor-rmat-row-aux (i j a n)
    (if (and (natp n) (> n 1) (natp j) (< j n))
        (cons (rdet-cofactor i j a n) (cofactor-rmat-row-aux i (1+ j) a n))
      ()))
  (defund cofactor-rmat-row (i a n) (cofactor-rmat-row-aux i 0 a n))
  
  (defun cofactor-rmat-aux (i a n)
    (if (and (natp n) (natp i) (< i n))
        (cons (cofactor-rmat-row i a n) (cofactor-rmat-aux (1+ i) a n))
      ()))
  (defund cofactor-rmat (a n) (cofactor-rmat-aux 0 a n))
\end{verbatim}
\end{small}
The {\it classical adjoint} of {\tt a} is the transpose of its cofactor matrix:

\begin{small}
\begin{verbatim}
  (defund adjoint-rmat (a n) (transpose-mat (cofactor-rmat a n)))
\end{verbatim}
\end{small}
The following is an equivalent formulation:

\begin{small}
\begin{verbatim}
  (defthmd cofactor-rmat-transpose
    (implies (and (rmatp a n n) (natp n) (> n 1))
             (equal (cofactor-rmat (transpose-mat a) n)
                    (adjoint-rmat a n))))
\end{verbatim}
\end{small}
Note that the dot product of {\tt (row i a)} with {\tt (cofactor-rmat-row i a n)} is a rearrangement of the sum {\tt (expand-rdet-row a i n)}:

\begin{small}
\begin{verbatim}
  (defthmd rdot-cofactor-rmat-row-expand-rdet-row
    (implies (and (rmatp a n n) (natp n) (> n 1) (natp i) (< i n))
             (equal (rdot (row i a) (cofactor-rmat-row i a n))
                    (expand-rdet-row a i n))))
\end{verbatim}
\end{small}
Combining this with {\tt expand-rdet-row-rdet}, we have the following expression for the determinant:

\begin{small}
\begin{verbatim}
  (defthmd rdot-cofactor-rmat-row-rdet
    (implies (and (rmatp a n n) (natp n) (> n 1) (natp i) (< i n))
             (equal (rdot (row i a) (cofactor-rmat-row i a n))
                    (rdet a n))))
\end{verbatim}
\end{small}
Next we consider the result of substituting {\tt (replace-row a i (row k a))} for {\tt a} in {\tt rdot-cofactor\-rmat-row-rdet}, where {\tt k} $\neq$ {\tt i}.
Since this matrix has two identical rows, its determinant is 0, and we have

\begin{small}
\begin{verbatim}
  (defthmd rdot-cofactor-rmat-row-rdet-0
    (implies (and (rmatp a n n) (natp n) (> n 1) (natp i) (< i n)
                  (natp k) (< k n) (not (= k i)))
             (equal (rdot (row k a) (cofactor-rmat-row i a n))
                    (r0))))
\end{verbatim}
\end{small}
Thus, we have the following for general {\tt k}:

\begin{small}
\begin{verbatim}
  (defthmd rdot-cofactor-rmat-row-rdelta
    (implies (and (rmatp a n n) (natp n) (> n 1) (natp i) (< i n) (natp k) (< k n))
             (equal (rdot (row k a) (cofactor-rmat-row i a n))
                    (r* (rdelta i k) (rdet a n)))))
\end{verbatim}
\end{small}
Since {\tt (cofactor-rmat-row i a n)} = {\tt (col i (adjoint-mat a n))}, this yields an expression for the {\tt n}$\times${\tt n} matrix product of a and its adjoint:

\begin{small}
\begin{verbatim}
  (defthmd rmat*-adjoint-rmat
    (implies (and (rmatp a n n) (natp n) (> n 1))
             (equal (rmat* a (adjoint-rmat a n))
                    (rmat-scalar-mul (rdet a n) (id-rmat n)))))
\end{verbatim}
\end{small}

In Part II, where we consider matrices with entries ranging over a field, we shall use this last equation in deriving a criterion for the
existence of a multiplicative inverse of a matrix.  We shall also apply the results of this subsection to a proof of Cramer's Rule for solving
a system of {\tt n} linear equations in {\tt n} unknowns.

\bibstyle{eptcs}
\bibliographystyle{eptcs}
\bibliography{linear1}

\begin{thebibliography}{10}
\providecommand{\bibitemdeclare}[2]{}
\providecommand{\surnamestart}{}
\providecommand{\surnameend}{}
\providecommand{\urlprefix}{Available at }
\providecommand{\url}[1]{\texttt{#1}}
\providecommand{\href}[2]{\texttt{#2}}
\providecommand{\urlalt}[2]{\href{#1}{#2}}
\providecommand{\doi}[1]{doi:\urlalt{https://doi.org/#1}{#1}}
\providecommand{\eprint}[1]{arXiv:\urlalt{https://arxiv.org/abs/#1}{#1}}
\providecommand{\bibinfo}[2]{#2}

\bibitemdeclare{book}{brown}
\bibitem{brown}
\bibinfo{author}{William \surnamestart Brown\surnameend}
  (\bibinfo{year}{1993}): \emph{\bibinfo{title}{Matrices over Commutative
  Rings}}.
\newblock \bibinfo{publisher}{M. Dekker}.

\bibitemdeclare{inproceedings}{cowles}
\bibitem{cowles}
\bibinfo{author}{Ruben \surnamestart Gamboa\surnameend}, \bibinfo{author}{John
  \surnamestart Cowles\surnameend} \& \bibinfo{author}{Jeff~Van \surnamestart
  Baalen\surnameend} (\bibinfo{year}{2003}): \emph{\bibinfo{title}{Using ACL2
  Arrays to Formalize Matrix Algebra}}.
\newblock In: {\slshape \bibinfo{booktitle}{ACL2 2003: 4th International
  Workshop on the ACL2 Theorem Prover and its Applications}},
  \bibinfo{address}{Boulder, Colorado}.

\bibitemdeclare{inproceedings}{hendrix}
\bibitem{hendrix}
\bibinfo{author}{Joe \surnamestart Hendrix\surnameend} (\bibinfo{year}{2003}):
  \emph{\bibinfo{title}{Matrices in ACL2}}.
\newblock In: {\slshape \bibinfo{booktitle}{ACL2 2003: 4th International
  Workshop on the ACL2 Theorem Prover and its Applications}},
  \bibinfo{address}{Boulder, Colorado}.

\bibitemdeclare{book}{hoffman}
\bibitem{hoffman}
\bibinfo{author}{Kenneth \surnamestart Hoffman\surnameend} \&
  \bibinfo{author}{Ray \surnamestart Kunze\surnameend} (\bibinfo{year}{1961}):
  \emph{\bibinfo{title}{Linear Algebra}}.
\newblock \bibinfo{publisher}{Allyn Prentice-Hall}.

\bibitemdeclare{book}{kolman}
\bibitem{kolman}
\bibinfo{author}{Bernard \surnamestart Kolman\surnameend}
  (\bibinfo{year}{1977}): \emph{\bibinfo{title}{Elementary Linear Algebra}},
  \bibinfo{edition}{2nd} edition.
\newblock \bibinfo{publisher}{MacMillan}.

\bibitemdeclare{book}{kwak}
\bibitem{kwak}
\bibinfo{author}{Jin~Ho \surnamestart Kwak\surnameend} \&
  \bibinfo{author}{Sungpyo \surnamestart Kong\surnameend}
  (\bibinfo{year}{1997}): \emph{\bibinfo{title}{Linear Algebra}}.
\newblock \bibinfo{publisher}{Birkh{\"a}user}. \doi{10.1007/978-1-4757-1200-1}.

\bibitemdeclare{inproceedings}{kwan1}
\bibitem{kwan1}
\bibinfo{author}{Carl \surnamestart Kwan\surnameend} \& \bibinfo{author}{Warren
  \surnamestart Hunt\surnameend} (\bibinfo{year}{2024}):
  \emph{\bibinfo{title}{Automatic Verification of Right-greedy Numerical Linear
  Algebra Algorithms}}.
\newblock In: {\slshape \bibinfo{booktitle}{Proceedings of the 24th Conference
  on Formal Methods in Computer-Aided Design (FMCAD 2024)}},
  \doi{10.34727/2024/isbn.978-3-85448-065-5}.

\bibitemdeclare{inproceedings}{kwan2}
\bibitem{kwan2}
\bibinfo{author}{Carl \surnamestart Kwan\surnameend} \& \bibinfo{author}{Warren
  \surnamestart Hunt\surnameend} (\bibinfo{year}{2024}):
  \emph{\bibinfo{title}{Formalizing the Cholesky Factorization Theorem}}.
\newblock In: {\slshape \bibinfo{booktitle}{Proceedings for the Fifteenth
  Conference on Interactive Theorem Proving (ITP 2024)}},
  \doi{10.4230/LIPIcs.ITP.2024.25}.

\bibitemdeclare{misc}{lean}
\bibitem{lean}
\emph{\bibinfo{title}{Maths in Lean: Linear Algebra}}.
\newblock \bibinfo{note}{Available at
  \burl{https://leanprover-community.github.io/theories/linear_algebra.html}}.

\bibitemdeclare{book}{roman}
\bibitem{roman}
\bibinfo{author}{Steven \surnamestart Roman\surnameend} (\bibinfo{year}{2005}):
  \emph{\bibinfo{title}{Advanced Linear Algebra}}, \bibinfo{edition}{2nd}
  edition.
\newblock \bibinfo{publisher}{Springer}, \doi{10.1007/978-1-4757-2178-2}.

\bibitemdeclare{inproceedings}{groups1}
\bibitem{groups1}
\bibinfo{author}{David~M. \surnamestart Russinoff\surnameend}
  (\bibinfo{year}{2022}): \emph{\bibinfo{title}{A Formalization of Finite Froup
  Theory}}.
\newblock In: {\slshape \bibinfo{booktitle}{ACL2 2022: 17th International
  Workshop on the ACL2 Theorem Prover and its Applications}},
  \bibinfo{address}{Austin, Texas}, \doi{10.4204/EPTCS.359.10}.

\bibitemdeclare{inproceedings}{groups2}
\bibitem{groups2}
\bibinfo{author}{David~M. \surnamestart Russinoff\surnameend}
  (\bibinfo{year}{2023}): \emph{\bibinfo{title}{A Formalization of Finite Froup
  Theory: Part II}}.
\newblock In: {\slshape \bibinfo{booktitle}{ACL2 2023: 18th International
  Workshop on the ACL2 Theorem Prover and its Applications}},
  \bibinfo{address}{Austin, Texas}, \doi{10.4204/EPTCS.393.4}.

\bibitemdeclare{inproceedings}{groups3}
\bibitem{groups3}
\bibinfo{author}{David~M. \surnamestart Russinoff\surnameend}
  (\bibinfo{year}{2023}): \emph{\bibinfo{title}{A Formalization of Finite Froup
  Theory: Part III}}.
\newblock In: {\slshape \bibinfo{booktitle}{ACL2 2023: 18th International
  Workshop on the ACL2 Theorem Prover and its Applications}},
  \bibinfo{address}{Austin, Texas}, \doi{10.4204/EPTCS.393.5}.

\bibitemdeclare{inproceedings}{linear2}
\bibitem{linear2}
\bibinfo{author}{David~M. \surnamestart Russinoff\surnameend}
  (\bibinfo{year}{2025}): \emph{\bibinfo{title}{A Formalization of Elementary
  Linear Algebra: Part II}}.
\newblock In: {\slshape \bibinfo{booktitle}{ACL2 2025: 19th International
  Workshop on the ACL2 Theorem Prover and its Applications}},
  \bibinfo{address}{Austin, Texas}.

\bibitemdeclare{incollection}{coq}
\bibitem{coq}
\bibinfo{author}{ZhengPu \surnamestart Shi\surnameend} \& \bibinfo{author}{Gang
  \surnamestart Chen\surnameend} (\bibinfo{year}{2022}):
  \emph{\bibinfo{title}{Integration of Multiple Formal Matrix Models in Coq}}.
\newblock In \bibinfo{editor}{Wei \surnamestart Dong\surnameend} \&
  \bibinfo{editor}{Jean-Pierre \surnamestart Talpin\surnameend}, editors:
  {\slshape \bibinfo{booktitle}{Dependable Software Engineering Theories,
  Tools, and Applications}}, \bibinfo{publisher}{Springer Nature Switzerland},
  \doi{10.1007/978-3-031-21213-0_11}.

\bibitemdeclare{inproceedings}{coq2}
\bibitem{coq2}
\bibinfo{author}{ZhengPu \surnamestart Shi\surnameend} \& \bibinfo{author}{Gang
  \surnamestart Chen\surnameend} (\bibinfo{year}{2024}):
  \emph{\bibinfo{title}{Formal Verification of Executable Matrix Inversion via
  Adjoint Matrix and Gaussian Elimination}}.
\newblock In: {\slshape \bibinfo{booktitle}{Proceedings of the 26th
  International Symposium on Principles and Practice of Declarative
  Programming}}, \doi{10.1145/3678232.3678242}.

\bibitemdeclare{incollection}{hol4}
\bibitem{hol4}
\bibinfo{author}{Zhiping \surnamestart Shi\surnameend}, \bibinfo{author}{Yan
  \surnamestart Zhang\surnameend}, \bibinfo{author}{Zhenke \surnamestart
  Liu\surnameend}, \bibinfo{author}{Ximan \surnamestart Kank\surnameend},
  \bibinfo{author}{Yong \surnamestart Guan\surnameend}, \bibinfo{author}{Jie
  \surnamestart Zhang\surnameend} \& \bibinfo{author}{Xiaoyu \surnamestart
  Song\surnameend} (\bibinfo{year}{2014}): \emph{\bibinfo{title}{Formalization
  of matrix theory in Hol4}}.
\newblock In: {\slshape \bibinfo{booktitle}{Advances in Mechanical
  Engineering}}, \bibinfo{volume}{6}, \doi{10.1155/2014/195276}.

\bibitemdeclare{incollection}{isabelle}
\bibitem{isabelle}
\bibinfo{author}{Christian \surnamestart Sternagel\surnameend} \&
  \bibinfo{author}{Rene \surnamestart Thiemann\surnameend}
  (\bibinfo{year}{2010}): \emph{\bibinfo{title}{Executable Matrix Operations on
  Matrices of Arbitrary Dimensions}}.
\newblock In: {\slshape \bibinfo{booktitle}{Archive of Formal Proofs}}.
\newblock \bibinfo{note}{Available at
  \burl{https://www.isa-afp.org/entries/Matrix.html}}.

\end{thebibliography}
\end{document}